\documentclass[11pt,a4paper]{article}

\usepackage{epsfig}
\usepackage{amssymb}
\usepackage{amsmath}

\begin{document}

\title{Strong field limit analysis of gravitational retro-lensing}
\author{Ernesto F. Eiroa$^{1,}$\thanks{
e-mail: eiroa@iafe.uba.ar} , and Diego F. Torres$^{2,}$\thanks{
e-mail: dtorres@igpp.ucllnl.org} \\
{\small $^1$ Instituto de Astronom\'{\i}a y F\'{\i}sica del Espacio, C.C.
67, Suc. 28, 1428, Buenos Aires, Argentina}\\
{\small $^2$ Lawrence Livermore National Laboratory, 7000 East Ave., L-413,
Livermore, CA 94550, USA}}
\maketitle
\date{}

\begin{abstract}
We present a complete treatment in the strong field limit of gravitational 
retro-lensing by a static spherically symmetric compact object having 
a photon sphere. The results are compared
with those corresponding to ordinary lensing in similar strong
field situations. As examples of application of the formalism, a
supermassive black hole at the galactic center and a stellar mass
black hole in the galactic halo are studied as retro-lenses, in both cases 
using Schwarzschild and Reissner-Nordstr\"{o}m geometries.
\end{abstract}



\section{Introduction}

Gravitational lensing is an important tool in astrophysics. Observable 
phenomena of lensing by stars and galaxies can be explained in the weak field 
approximation, that is to say, keeping only the first order term in the 
expansion of the deflection angle \cite{schne}. But when the lens is a black 
hole, a full general relativistic treatment is in order.\\

Recently, several articles studying strong field lensing appeared in the
literature. Virbhadra \& Ellis \cite{virbha1} numerically analyzed  the
situation where the lens is a Schwarzschild black hole placed in the center
of the Galaxy. They obtained the lens equation with an asymptotically flat
metric and found that, besides the primary and secondary images, two infinite
sets of faint relativistic images are formed by photons that make complete
turns (in both directions of rotation) around the black hole before reaching
the observer. Fritelli, Kling \& Newman
\cite{fritelli} found an exact lens equation without any reference to a
background metric and compared their results with those of Virbhadra \& Ellis.
Bozza et al \cite{bozza1} developed an approximate analytical method 
(generally called, as well,
the strong field limit) by which they obtained the positions and
magnifications of the relativistic images for the Schwarzschild black hole.
Subsequently, Eiroa, Romero \& Torres \cite{eiroto} applied 
the strong field limit to the Reissner-Nordstr\"{o}m black hole and
discussed the possibility of detection of the relativistic images in the
next few years. Bozza \cite{bozza2} extended the strong field limit to
any static spherically symmetric lens and analyzed the case of a charged
Brans-Dicke black hole. Bhadra \cite{bhadra} studied the
Gibbons-Maeda-Grafinkle-Horowitz-Strominger charged black hole of string
theory in the strong field limit. Virbhadra \& Ellis numerically investigated
the lensing by naked singularities \cite{virbha2}. Bozza \& Mancini
\cite{bozza3} used the strong field limit to study the time delay between
different relativistic images, showing that different types of black holes
are characterized by different time delays.
Spinning black holes are more difficult to tackle. Bozza \cite{bozza4}
analyzed the case of quasi equatorial lensing by rotating black holes and
V\'{a}zquez \& Esteban \cite{vazquez} studied  the Kerr black hole where the
observer and source have arbitrary inclinations. Many of these recent 
advances (if not all) deal with theoretical implications of General 
Relativity in a regime which is well beyond our
current technological capabilities for detection. Henceforth, the emphasis 
of these studies is not as much in predicting new phenomena that could 
be tested in a short period of time as it is in gaining
insight on how gravitational lensing in particular, and gravity in general, 
behave in the strong field regimes.
\\

In ordinary lensing situations, the lens is placed between the source and the
observer. But if the lens is a compact object with a photon sphere and the
observer is placed  between the source and the lens, it leads to the
formation of images with deflection angles closer to odd multiples of $\pi$.
The observer sees the images in front and the source behind.
This situation is called retro-lensing and was studied for the first time by
Holtz \& Wheeler \cite{holtz}, who analyzed only the two stronger images for a
black hole placed in the galactic bulge with the sun as source. They also
proposed retro-lensing as a new mechanism for searching black
holes. A similar idea was suggested in the context of defocusing
gravitational lensing by Capozziello et al \cite{capoz}. De Paolis et al
\cite{depaolis} recently analyzed the retro-lensing scenario for a bright
star close to the massive black hole at the galactic center.\\

In this paper we give a complete treatment of retro-lensing by a static,
spherically symmetric lens, using the strong field limit. In Section 2 we
obtain the lens equation for retro-lensing and the deflection angle
in the strong field limit. In Section 3, the position and
magnification of the images are calculated. In Section 4, the results are
compared with those for ordinary lensing and astrophysical
examples are given in Section 5. Final concluding remarks are given in 
Section 6. Throughout the paper we use units such that $G=c=1$.

\section{Lens equation and deflection angle}

\begin{figure}[t!]
\vspace{-4cm}
\begin{center}
\includegraphics[width=15cm]{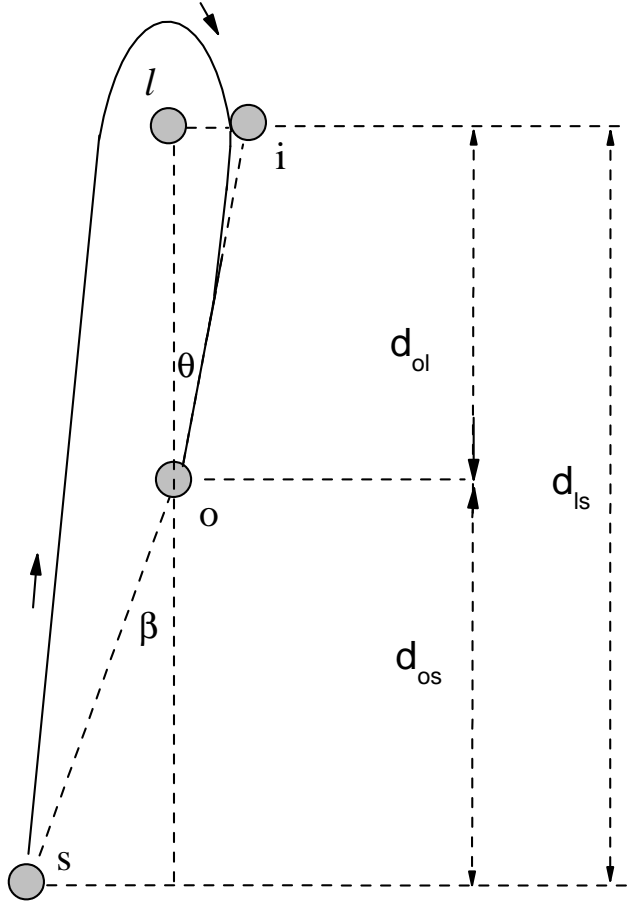}
\end{center}
\vspace{-3cm}
\caption{Lens diagram. The observer (o), the lens (l), the source (s) and
the projection on the lens plane of the first relativistic image (i) are
shown. The angular position of the source is $\beta$, of the image is
$\theta $, and $d_{ol}$, $d_{os}$, $d_{ls}$ are, respectively,
the observer-lens, the observer-source and the lens-source distances. Two
infinite sets of relativistic images (with both directions of rotation around
the lens) are obtained with deflection angles close to odd multiples of
$\pi $.}
\label{f1}
\end{figure}

Consider an observer (o) placed between a point source of light (s)
and a strong field object with a photon sphere (e.g. a black hole), which we
will call the lens (l). The line joining the observer and the lens
define the optical axis. The background space-time is considered
asymptotically flat, with the observer and the source immersed in the flat
region. We call $\beta $ the angular position of the source
 and $\theta $ the angular position of the images (i) as seen by the observer.
The lens situation is shown in Fig. \ref{f1}. We can take $\beta >0$ without
loosing generality.
The lens equation for retro-lensing is slightly different from ordinary
lensing (i.e. when the lens lies between the source and the observer), due to
the different relative positions of the lens, source and observer:
\begin{equation}
\tan \beta =\tan \theta -\frac{d_{os}}{d_{ls}}\left[ \tan (\alpha -\theta)
+\tan \theta \right],
\label{le1}
\end{equation}
where $d_{os}=D_{os}/2M$ and $d_{ls}=D_{ls}/2M$ are, respectively, the
observer-source and the lens-source distances in units of the Schwarzschild
radius, and $\alpha $ is the deflection angle. In lensing situations where
the objects are highly aligned, the angles $\beta $
and $\theta $ are small and $\alpha $ is closer to an odd multiple of $\pi $.
For the images at the opposite side of the source (see Fig. \ref{f1}), it can
be written as $\alpha =(2n-1)\pi +\Delta \alpha _{n}$, with $n\in
\mathbb{N}$ and $0<\Delta \alpha _{n}\ll 1$. Then, the lens equation takes the
form
\begin{equation}
\beta =\theta -\frac{d_{os}}{d_{ls}}\Delta \alpha _{n}.
\label{le2}
\end{equation}
As in the case of ordinary strong field lensing, two infinite sets of
relativistic images are formed. To obtain the other set of images (at
the same side of the source) we should
take $\alpha =-(2n-1)\pi -\Delta \alpha _{n}$, so $\Delta \alpha _{n}$ must
be replaced by $-\Delta \alpha _{n}$ in Eq. (\ref{le2}). In the case of
perfect alignment, an infinite series of concentric Einstein rings are
obtained.\\

To calculate the deflection angle, consider an static
spherically symmetric metric
\begin{equation}
ds^{2}=-f(x)dt^{2}+g(x)dx^{2}+h(x)d\Omega ^{2},
\label{met}
\end{equation}
where $x=r/2M$ is the radial coordinate in units of the Schwarzschild
radius. We require this metric to have a photon sphere, which radius $x_{ps}$
is given by the greater positive solution of the equation:
\begin{equation}
\frac{h^{\prime }(x)}{h(x)}=\frac{f^{\prime }(x)}{f(x)}.
\label{ps}
\end{equation}
where the prime means derivative with respect to $x$.\\

The deflection angle corresponding to the images situated at the opposite
side of the source as a function of the closest distance of approach $x_{0}$
is given by \cite{weinberg}
\begin{equation}
\alpha (x_{0})=I(x_{0})-\pi ,
\label{alp1}
\end{equation}
with
\begin{equation}
I(x_{0})=\int_{x_{0}}^{\infty }2\left( \frac {g(x)}{h(x)}\right)^{1/2}
\left( \frac {h(x)f(x_{0})}{h(x_{0})f(x)}-1\right)^{-1/2}dx.
\label{i1}
\end{equation}

The impact parameter $b$ (in units of the Schwarzschild radius) is
\cite{weinberg}
\begin{equation}
b(x_{0})=\left( \frac{h(x_{0})}{f(x_{0})}\right) ^{1/2}.
\label{b1}
\end{equation}
From the lens geometry we also have that
\begin{equation}
b(x_{0})=d_{ol}\sin \theta \approx d_{ol}\theta ,
\label{b2}
\end{equation}
with $d_{ol}=D_{ol}/2M$ the observer-lens distance in units of the
Schwarzschild radius.\\

Bozza \cite{bozza2} has shown that for a spherically symmetric lens with a
photon sphere, light rays with impact parameter $b$
close to the photon sphere have a  deflection angle that can be approximated by
\begin{equation}
\alpha =-a_{1}\ln (b-b_{ps})+a_{2}+ O(b-b_{ps}),
\label{alp2}
\end{equation}
where $b_{ps}=b(x_{ps})$, and $a_{1,2}$ are constants which depend only on
the type of lens. Provided the metric of the compact object that acts as a
lens, the coefficients $a_{1,2}$ can be obtained in a laborious but
straightforward way using the formalism developed in \cite{bozza2}. The
method consists essentially in separating the integral of Eq. (\ref{i1}) in a
divergent part which gives the logarithmic term in Eq. (\ref{alp2}), and a
regular part which gives the constant $a_{2}$. This approximation, named the
strong field limit, gives very accurate results. For more details see
\cite{bozza2}. \\

Another way of calculating the constants $a_{1,2}$ is following a similar
approach to that used by Eiroa, Romero \& Torres \cite{eiroto}:
\begin{equation}
a_{1}=\lim_{x_{0}\rightarrow x_{ps}}\frac{-[b(x_{0})-b(x_{ps})]
I^{\prime }(x_{0})}{b^{\prime }(x_{0})},
\label{alp3}
\end{equation}
and
\begin{equation}
a_{2}=\ln \left\{ \lim_{x_{0}\rightarrow x_{ps}}\left[ e^{I(x_{0})-\pi}
 [b(x_{0})-b(x_{ps})]^{a_{1}} \right] \right\} ,
\label{alp4}
\end{equation}
with the prime now meaning the derivative with respect to $x_{0}$. These
limits can be evaluated numerically with standard software.\\

The values of the coefficients $a_{1}$ and $a_{2}$ for Schwarzschild and
Reissner-Nordstr\"{o}m lens geometries are given in the Appendix.

\section{Position and magnification of the images}

Inverting Eq. (\ref{alp2}) to obtain the impact parameter
\begin{equation}
b=b_{ps}+\exp \left( \frac{a_{2}-\alpha }{a_{1}} \right),
\label{b3}
\end{equation}
and using that $b=d_{ol}\theta $, we have that the position of any image is
\begin{equation}
\theta =\frac {b_{ps}}{d_{ol}}+\frac {1}{d_{ol}}\exp \left(
\frac{a_{2}-\alpha }{a_{1}} \right).
\label{th1}
\end{equation}
Taylor expanding to first order around $\alpha =(2n-1)\pi $, we can
approximate the angular position of the $n$-th image by
\begin{equation}
\theta _{n}=\theta ^{0}_{n}-\frac {\xi _{n}}{d_{ol}}\Delta \alpha _{n},
\label{th2}
\end{equation}
with
\begin{equation}
\theta ^{0}_{n}=\frac {b_{ps}}{d_{ol}}+\frac {1}{d_{ol}}\exp \left[
\frac{a_{2}-(2n-1)\pi }{a_{1}} \right],
\label{th3}
\end{equation}
and
\begin{equation}
\xi_{n}=\frac {1}{a_{1}}\exp \left[ \frac{a_{2}-(2n-1)\pi }{a_{1}} \right].
\label{th4}
\end{equation}
Then from Eq. (\ref{th2}) we obtain
\begin{equation}
\Delta \alpha _{n}=-\frac {\theta _{n}-\theta ^{0}_{n}}{\xi _{n}}d_{ol},
\label{alp5}
\end{equation}
and replacing it in the lens equation
\begin{equation}
\beta =\theta _{n}+\frac{d_{os}d_{ol}}{d_{ls}\xi _{n}}
(\theta _{n}-\theta ^{0}_{n}).
\label{le3}
\end{equation}
The last equation can be written in the form
\begin{equation}
\beta -\theta ^{0}_{n}=\left( 1+\frac{d_{os}d_{ol}}{d_{ls}\xi _{n}}\right)
(\theta _{n}-\theta ^{0}_{n}),
\label{le4}
\end{equation}
and as the second term in the parentheses is much greater than one, we have
that
\begin{equation}
\beta -\theta ^{0}_{n}\approx \frac {d_{os}d_{ol}}{d_{ls}\xi _{n}}
(\theta _{n}-\theta ^{0}_{n}).
\label{le5}
\end{equation}
We finally obtain that the angular positions of the images are
\begin{equation}
\theta _{n}=\theta ^{0}_{n}+\frac {d_{ls}\xi _{n}}{d_{os}d_{ol}}(\beta -
\theta ^{0}_{n}).
\label{th5a}
\end{equation}
The second term in Eq. (\ref{th5a}) is a small correction on $\theta ^{0}_{n}$,
so all images lie very close to $\theta ^{0}_{n}$. With a similar treatment,
the other set of relativistic images have positions given by
\begin{equation}
\theta _{n}=-\theta ^{0}_{n}+\frac {d_{ls}\xi _{n}}{d_{os}d_{ol}}(\beta +
\theta ^{0}_{n}).
\label{th5b}
\end{equation}
When $\beta =0$ an infinite sequence of Einstein rings is formed,  with 
angular radius
\begin{equation}
\theta ^{E}_{n}=\theta ^{0}_{n}-\frac {d_{ls}\xi _{n}}{d_{os}d_{ol}}
\theta ^{0}_{n}.
\label{th6}
\end{equation}
\\
Gravitational lensing conserves surface brightness, so the ratio of the solid
angles subtended by the image and the source gives the amplification of the
$n$-th image:
\begin{equation}
\mu _{n}=\left| \frac{\sin \beta }{\sin \theta _{n}}
\frac{d\beta }{d\theta _{n}}\right|^{-1} \approx
\left| \frac{\beta }{\theta _{n}} \frac{d\beta }{d\theta _{n}}\right|^{-1}.
\label{mu1}
\end{equation}
Using Eq. (\ref{th5a}) we have that
\begin{equation}
\mu _{n}=\frac{1}{\beta}\left[ \theta ^{0}_{n}+
\frac {d_{ls}\xi _{n}}{d_{os}d_{ol}}(\beta - \theta ^{0}_{n})\right]
\frac {d_{ls}\xi _{n}}{d_{os}d_{ol}},
\label{mu2}
\end{equation}
which can be approximated to first order in
$ {d_{ls}}/{d_{os}d_{ol}}\ll 1$ by
\begin{equation}
\mu _{n}=\frac{1}{\beta}\frac {d_{ls}}{d_{os}d_{ol}}
\theta ^{0}_{n}\xi _{n}.
\label{mu3}
\end{equation}
The same result is obtained for the other set of images. The first
relativistic image is the brightest one, and the magnifications
decrease exponentially with $n$. \\

The total magnification, considering both sets of images, is
$\mu =2\sum\limits_{n=1}^{\infty }\mu _{n}$, which can be written as
\begin{equation}
\mu =\frac {2}{\beta}\frac {d_{ls}a_{3}}{d_{os}d_{ol}^{2}a_{1}},
\label{mu4}
\end{equation}
where
\begin{equation}
a_{3}=b_{ps}\frac {\exp \left( \frac {a_{2}-\pi }{a_{1}}\right) }
{1-\exp \left( -\frac {2\pi }{a_{1}}\right) }
+\frac {\exp \left( \frac {2a_{2}-2\pi }{a_{1}}\right) }
{1-\exp \left(-\frac {4\pi }{a_{1}}\right) }.
\label{mu5}
\end{equation}
\\
\underline{Extended sources}
\\

If the source is extended, we have to integrate over its luminosity profile to
obtain the magnification:
\begin{equation}
\mu =\frac{\int\!\!\int_{S}\mathcal{I}\mu _{p}dS}{\int\!\!\int_{S}
\mathcal{I}dS}, \label{ext1}
\end{equation}
where $\mathcal{I}$ is the surface intensity distribution of the source and
$\mu _{p}$ is the magnification corresponding to each point of the source.
If the source is uniform, and we use polar coordinates $(R,\varphi )$ in the
source plane, with $R=0$ in the optical axis, and we consider the source as
a disk $D(R_{c},R_{s})$ of radius $R_{s}$ centered in $R_{c}$, the last
equation takes the form
\begin{equation}
\mu = \frac{\int\!\!\int_{D(R_{c},R_{s})}\mu _{p }RdRd\varphi }
{\pi R_{s}^{2}}, \label{ext2}
\end{equation}
and, using that $\beta =R/D_{os}$ is the angular position of each point
of the source, we have that
\begin{equation}
\mu =\frac{\int\!\!\int_{D(\beta _{c},\beta _{s})}\mu _{p}
\beta d\beta d\varphi }{\pi \beta _{s}^{2}},
\label{ext3}
\end{equation}
where $D(\beta _{c},\beta _{s})$ is the disk with angular radius $\beta _{s}$
centered in $\beta _{c}$. Then
\begin{equation}
\mu _{n}=\frac{I}{\pi \beta _{s}^{2}}\frac {d_{ls}}{d_{os}d_{ol}}
\theta ^{0}_{n}\xi _{n},
\label{ext4}
\end{equation}
and
\begin{equation}
\mu =\frac{2I}{\pi \beta _{s}^{2}}\frac{d_{ls}a_{3}}{d_{os}d_{ol}^{2}a_{1}},
\label{ext5}
\end{equation}
with $I=\int\!\!\int_{D(\beta _{c},\beta _{s})} d\beta d\varphi$. This
integral can be calculated in terms of elliptic functions \cite{bozza1}:
\begin{equation}
I=2Sign[\beta _{s}-\beta _{c}]
\left[ (\beta _{s}-\beta _{c})E\left( \frac{\pi }{2},
\frac{-4\beta _{s}\beta _{c}}{(\beta _{s}-\beta _{c})^{2}}\right)
+(\beta _{s}+\beta _{c})F\left( \frac{\pi }{2},
\frac{-4\beta _{s}\beta _{c}}{(\beta _{s}-\beta _{c})^{2}}\right) \right] ,
\label{ext6}
\end{equation}
where
\begin{equation}
F(\phi _{0},\lambda )=\int_{0}^{\phi _{0}}\left( 1-\lambda \sin ^{2}\phi
\right) ^{-\frac{1}{2}}d\phi ,  \label{ext7}
\end{equation}
\begin{equation}
E(\phi _{0},\lambda )=\int_{0}^{\phi _{0}}\left( 1-\lambda \sin ^{2}\phi
\right) ^{\frac{1}{2}}d\phi ,  \label{ext8}
\end{equation}
are respectively, elliptic integrals of the the first and second kind with
the arguments $\phi _{0}$ and $\lambda $.

\section{Comparison with ordinary lensing}

In ordinary lensing the angular positions of the two sets of relativistic
images, for a point source, are given by
\begin{equation}
\theta _{n}=\theta ^{0}_{n}+\frac {d_{os}\xi _{n}}{d_{ls}d_{ol}}(\beta -
\theta ^{0}_{n}),
\label{ol1a}
\end{equation}
and
\begin{equation}
\theta _{n}=-\theta ^{0}_{n}+\frac {d_{os}\xi _{n}}{d_{ls}d_{ol}}(\beta +
\theta ^{0}_{n}),
\label{ol1b}
\end{equation}
with
\begin{equation}
\theta ^{0}_{n}=\frac {b_{ps}}{d_{ol}}+\frac {1}{d_{ol}}\exp \left(
\frac{a_{2}-2n\pi }{a_{1}} \right),
\label{ol3}
\end{equation}
and
\begin{equation}
\xi_{n}=\frac {1}{a_{1}}\exp \left( \frac{a_{2}-2n\pi }{a_{1}} \right).
\label{ol4}
\end{equation}
We see that for both cases (ordinary and retro-lensing), the images are formed
close to $\theta ^{0}_{n}$, which has values of the order of the angular
radius of the photon sphere of the lens. From Eqs. (\ref{th3}) and (\ref{ol3})
we observe that the images for retro-lensing lies a bit farther of the photon
sphere than in the ordinary case.\\

The magnifications in ordinary lensing, for a point source, of the
relativistic images are (both sets):
\begin{equation}
\mu _{n}=\frac{1}{\beta}\frac {d_{os}}{d_{ls}d_{ol}}
\theta ^{0}_{n}\xi _{n},
\label{ol2}
\end{equation}
and the total magnification, summing the magnifications of  both sets of
relativistic images, is
\begin{equation}
\mu =\frac {2}{\beta}\frac {d_{os}a_{3}}{d_{ls}d_{ol}^{2}a_{1}},
\label{ol5}
\end{equation}
where
\begin{equation}
a_{3}=b_{ps}\frac {\exp \left( \frac {a_{2}-2\pi }{a_{1}}\right) }
{1-\exp \left( -\frac {2\pi }{a_{1}}\right) }
+\frac {\exp \left( \frac {2a_{2}-4\pi }{a_{1}}\right) }
{1-\exp \left(-\frac {4\pi }{a_{1}}\right) }.
\label{ol6}
\end{equation}
Comparing Eqs. (\ref{mu5}) and (\ref{ol6}) we see that the first term of
$a_{3}$ in retro-lensing is greater by a factor
$e ^{\pi/a_{1}}$ and the second by a factor $e ^{2\pi/a_{1}}$ with respect
to the ordinary case. Then, taking by example the lensing by a Schwarzschild
black hole, and considering a situation where the factor involving the
distances and the source angular position is the same, the magnification
of the relativistic images in retro-lensing is greater by a factor about 24
compared with ordinary lensing.\\

In addition, with ordinary lensing, besides the two set of relativistic 
images, two weak field images with small deflection angles are produced by 
the lens. These weak field images do not appear in retro-lensing, and it is 
the main difference in both cases. The weak field images are orders of 
magnitude stronger than the relativistic ones \cite{bozza1, eiroto} and lie 
very close to them. This makes detecting the relativistic images in
ordinary lensing a very difficult task. In the case of retro-lensing, 
instead, the faint relativistic images are stronger and relatively easier to 
identify due to the absence of the brighter weak field images.

\section{Examples}

In this Section we apply the formalism to the Schwarzschild and
Reissner-Nordstr\"{o}m lens geometries. Two cases are analyzed: a 
super-massive black hole situated at the galactic center and a stellar mass 
black hole in the galactic halo. In both cases a nearby star is taken as 
source.\\

There is strong evidence of the existence of a super-massive black hole 
at the center of our galaxy
\cite{richstone}. We take a $M \sim 2.8\times 10^{6}M_{\odot}$ black hole as 
an fiducial lens at a distance $D_{ol}=8.5$ kpc. A
star with radius $R=R_{\odot}$ situated at $D_{os}=50$ pc is assumed as the 
source. In Table \ref{table1} the first (outer) Einstein ring angular position
 ($\theta_{1}^{E}$) and the limiting angular value of the Einstein rings
($\theta_{\infty }^{E}$) are given for the Schwarzschild and
Reissner-Nordstr\"{o}m black hole lenses. The images lie very close to the
Einstein angular radii. For extended sources, when there is complete
alignment, instead of a pair of images for each $n$ (one on each side), an
annular shaped image is formed, with magnification $2\mu _{n}$. The
magnification of the first relativistic image and the total magnification,
for perfect alignment, are also shown in Table \ref{table1} .\\

\begin{table}[h]
\begin{center}
\caption{Black hole in the galactic center as a gravitational retro-lens for
a star situated at 50 pc from the Earth. For several values of the
charge $q$ (in units of the Schwarzschild radius), the angular radii of the
photon sphere ($\theta_{ps}$), the first Einstein
ring ($\theta_{1}^{E}$), and the limiting value of the Einstein rings
($\theta_{\infty }^{E}$), are given. The magnifications, for perfect alignment
and taking the source as extended with radius $R_{\odot }$, of the first
image is $2\mu _{1}$, whereas the total magnification is $\mu $.}
\begin{tabular}{ccccccc}\hline
$\left| q\right| $ & $0$ & $0.1$ & $0.2$ & $0.3$ & $0.4$ & $0.5$\\ \hline
$\theta _{ps}  (\mu\textrm{arcsec})$ & $9.755$ & $9.667$ & $9.395$ & $8.899$ &
$8.080$ & $6.503$ \\
$\theta _{1}^{E} (\mu\textrm{arcsec})$ & $17.39$ & $17.28$ & $16.95$ &
$16.37$ & $15.53$ & $15.04$ \\
$\theta _{\infty }^{E} (\mu\textrm{arcsec})$ & $16.90$ & $16.78$ & $16.43$ &
$15.80$ & $14.78$ & $13.01$\\
$2\mu _{1} (\times 10^{-10})$ & $3.032$ &
$3.031$ & $3.049$ & $3.164$ & $3.673$ & $7.705$\\
$\mu (\times 10^{-10})$ & $3.037$ & $3.037$ &
$3.055$ & $3.172$ & $3.687$ & $7.784$\\\hline
\end{tabular}
\label{table1}
\end{center}
\end{table}

Low mass black holes are common in our galaxy
\cite{algol}. Let us now consider a stellar mass black hole with mass
$M=7M_{\odot}$ in the galactic halo with $D_{ol}=4$ kpc as retro-lens and a
star situated at $D_{os}=50$ pc with radius $R=R_{\odot}$ as source. The
results are given in Table \ref{table2}.\\

\begin{table}[h]
\begin{center}
\caption{Same as Table \ref{table1} for a black hole in the galactic halo as
a gravitational retro-lens with a star situated at 50 pc from the Earth as
source. }
\begin{tabular}{ccccccc}\hline
$\left| q\right| $ & $0$ & $0.1$ & $0.2$ & $0.3$ & $0.4$ & $0.5$\\ \hline
$\theta _{ps}  (\times 10^{-5} \mu\textrm{arcsec})$ & $5.182 $ &
$5.136$ & $ 4.991 $ & $4.728 $ & $4.292$ & $3.455$ \\
$\theta _{1}^{E} (\times 10^{-5} \mu\textrm{arcsec})$ & $9.236$ &
$9.178$ & $9.002$ & $8.696$ & $8.249$  & $ 7.990$ \\
$\theta _{\infty }^{E} (\times 10^{-5} \mu\textrm{arcsec})$ & $8.976$ &
$8.916$ & $8.729$ & $8.393$ & $7.853$ & $6.910$\\
$2\mu _{1} (\times 10^{-21})$ & $4.053$ &
$4.052$ & $4.076$ & $4.230$ & $4.911$ & $10.30$\\
$\mu (\times 10^{-21})$ & $4.060$ & $4.060$ &
$4.084$ & $4.240$ & $4.928$ & $10.41$\\\hline
\end{tabular}
\label{table2}
\end{center}
\end{table}

In both examples, the dependence of the angular positions and
magnifications of the images with the charge is weak, obtaining important
changes only for extreme values of charge.\\

As the images lie close to the photon sphere, the angular positions of them
take very small values, what increase with larger lens mass and smaller
observer-lens distance. For the supermassive black hole at the galactic
center, they are of the order of $10$ $\mu $arcsec, and for the stellar mass
black hole in the galactic halo they are about $10^{-5}$ $\mu $arcsec. 
Something similar happens with the magnifications, which also grow with 
greater lens mass and smaller observer-lens distance. For the supermassive 
black hole at the galactic center the magnifications are of order $10^{-10}$ 
and for the low mass one in the galactic halo of order $10^{-21}$.
\\

As long as $d_{os}\ll d_{ol}$, the angular positions and amplifications of
the images are not very sensitive to the value of $d_{os}$. So if we take the
Sun or a star at 100 pc from the Earth as source, the order of magnitude of
the results shown in Tables \ref{table1} and \ref{table2} does not change.
As the images appear highly demagnified, to improve the possibility of
observing them, strong nearby sources with nearly perfect alignment are
needed.\\

In all cases, the first image is the strongest one, carrying about 99\% of 
the total luminosity.

\section{Concluding Remarks}

We have presented the strong field limit formalism for gravitational
retrolensing. We have found the positions and magnifications of the 
relativistic images within the strong field limit, and studied some cases 
which might present particular interest and be a benchmark point for 
comparison with the normal lensing situation.
As in the case of ordinary lensing, the relativistic images are faint. 
However, if we compare the magnification of  retro-lensed relativistic images 
with those of the ordinary strong field lensing, we find that the former are 
significanlty stronger. Additionally, as no brighter weak field images appear 
in retro-lensing, it would be easier to detect them. The angular separation of
the relativistic images is of course beyond the angular resolution of current 
or foreseeable technologies, and not only angular resolution but also 
sensitivity improvements are needed to detect them (see the discussion in 
Eiroa, Romero \& Torres \cite{eiroto}). The image separations and 
magnifications present  values that make the observational identification of 
these relativistic  images a challenge for the future: we are not aware of 
any current research, in any wavelength, which would allow to resolve the 
retro-lensing images in the next years.
In the case of charged black holes, it is only for extreme values of
charge that important changes are obtained.

\section*{Acknowledgements}

This work has been partially supported by UBA (UBACYT X-143, EFE).
The work of DFT was performed under the auspices of the US DOE
(NNSA), by UC's LLNL under contract No. W-7405-Eng-48.

\section*{Appendix: Values of the coefficients $a_{1}$ and $a_{2}$}

\underline{Schwarzschild black hole}\\

The metric functions are
\begin{equation}
f(x)=g^{-1}(x)=1-\frac{1}{x}, \hspace{1cm}h(x)=x^{2},
\label{sch1}
\end{equation}
then
\begin{equation}
x_{ps}=\frac{3}{2},
\label{sch2}
\end{equation}
\begin{equation}
b(x_{0})=x_{0}\left( 1-\frac{1}{x_{0}}\right) ^{-1/2},
\hspace{1cm} b_{ps}=\frac{3\sqrt{3}}{2}.
\label{sch3}
\end{equation}
The coefficients can be calculated exactly \cite{bozza1,bozza2}
\begin{equation}
a_{1}=1,
\label{sch4}
\end{equation}
\begin{equation}
a_{2}=\ln [324(7\sqrt{3}-12)]-\pi .
\label{sch5}
\end{equation}
\\
\underline{Reissner-Nordstr\"{o}m black hole}\\

In this case
\begin{equation}
f(x)=g^{-1}(x)=1-\frac{1}{x}+\frac{q^{2}}{x^{2}}, \hspace{1cm}h(x)=x^{2},
\label{rn1}
\end{equation}
where $q=Q/2M$ is the electric charge in units of the Schwarzschild radius.
Then
\begin{equation}
x_{ps}=\frac{3}{4}\left[ 1+\left( 1-\frac{32}{9}q^{2}\right)^{1/2}  \right],
\label{rn2}
\end{equation}
\begin{equation}
b(x_{0})=x_{0}\left( 1-\frac{1}{x_{0}}+\frac{q^{2}}{x^{2}}\right) ^{-1/2},
\label{rn3a}
\end{equation}
\begin{equation}
b_{ps}=\frac{(3+\sqrt{9-32q^{2}})^{2}}
{4\sqrt{2}\sqrt{3-8q^{2}+\sqrt{9-32q^{2}}}}.
\label{rn3b}
\end{equation}
The coefficient $a_{1}$ can be calculated exactly \cite{bozza2}
\begin{equation}
a_{1}=\frac{x_{ps}\sqrt{x_{ps}-2q^{2}}}{\sqrt{(3-x_{ps})x^{2}_{ps}
-9q^{2}x_{ps}+8q^{4}}},
\label{rn4}
\end{equation}
and the coefficient $a_{2}$ can be approximated \cite{bozza2} by
\begin{equation}
a_{2}=c_{1}+c_{2}+a_{1}\ln b_{ps}-\pi
\label{rn5}
\end{equation}
with
\begin{equation}
c_{1}=2\ln [6(2-\sqrt {3})]+\frac{8}{9}\left\{ \sqrt {3}-4+\ln [6(2-\sqrt {3})]
\right\} q^{2}+O(q^{4}),
\label{rn5a}
\end{equation}
\begin{equation}
c_{2}=a_{1}\ln \left[ 2(x_{ps}-q^{2})^{2}\frac{(3-x_{ps})x^{2}_{ps}-9q^{2}
x_{ps}+8q^{4}}{(x_{ps}-2q^{2})^{3}(x^{2}_{ps}-x_{ps}+q^{2})}\right].
\label{rn5b}
\end{equation}



\begin{thebibliography}{99}

\bibitem{schne} P. Schneider, J. Ehlers and E.E. Falco, Gravitational Lenses
(Springer-Verlag, Berlin, 1992)

\bibitem{virbha1}  K.S. Virbhadra and G.F.R. Ellis, Phys. Rev. D \textbf{62},
084003 (2000).

\bibitem{fritelli}  S. Frittelli, T.P. Kling and E.T. Newman, Phys. Rev. D
\textbf{61}, 064021 (2000).

\bibitem{bozza1}  V. Bozza, S. Capozzielo, G. Iovane and G. Scarpetta, Gen.
Rel. Grav. \textbf{33}, 1535 (2001).

\bibitem{eiroto}  E.F. Eiroa, G.E. Romero and D.F. Torres, Phys. Rev. D
\textbf{66}, 024010 (2002).

\bibitem{bozza2}  V. Bozza, Phys. Rev. D  \textbf{66}, 103001 (2002).

\bibitem{bhadra}  A. Bhadra, Phys. Rev. D  \textbf{67}, 103009 (2003).

\bibitem{virbha2}  K.S. Virbhadra and G.F.R. Ellis, Phys. Rev. D \textbf{65},
103004 (2002).

\bibitem{bozza3}  V. Bozza and L. Mancini, arXiv:gr-qc/0305007 (2003).

\bibitem{bozza4}  V. Bozza, Phys. Rev. D  \textbf{67}, 103006 (2003).


\bibitem{vazquez} S.E. V\'{a}zquez and E.P. Esteban, arXiv:gr-qc/0308023
(2003).

\bibitem{holtz}  D.E. Holtz and J.A. Wheeler, Astrophys. J. \textbf{578}, 330
(2002).

\bibitem{capoz} S. Capozziello, R. de Ritis, V.I. Mank'o, A.A. Marino and
G. Marmo, Phys. Scripta \textbf{56}, 212 (1997).

\bibitem{depaolis}  F. De Paolis, A. Geralico, G. Ingrosso and A.A. Nucita,
arXiv:astro-ph/0307493 (2003).


\bibitem{weinberg}  S. Weinberg, Gravitation and Cosmology: Principles and
Applications of the General Theory of Relativity (Wiley, New York, 1972).

\bibitem{richstone}  R. Sch\"{o}del et al., Nature {\bf 419}, 694  (2002)


\bibitem{algol} E. Algol and M. Kamionkowski, Mon. Not. Roy. Astron. Soc.
\textbf{334}, 553 (2002).


\end{thebibliography}
\end{document}